\documentclass[showpacs,twocolumn,preprintnumbers,amsmath,amssymb]{revtex4-1}

\usepackage{graphicx}
\usepackage{dcolumn}
\usepackage{bm}
\usepackage{color}
\usepackage{amsmath}

\usepackage[T1]{fontenc}
\usepackage{mathptmx}

\begin{document}
\title{On-grating graphene surface plasmons enabling spatial differentiation in terahertz region}

\author{Yisheng Fang$^{1}$, Yijie Lou$^{1}$, Zhichao Ruan$^{1,2,3}$ }
\email[]{zhichao@zju.edu.cn}
\affiliation{$^1$ Department of Physics, Zhejiang University, Hangzhou 310027, China \\
$^2$ State Key Laboratory of Modern Optical Instrumentation, Zhejiang University, Hangzhou 310027, China \\
$^3$ College of Optical Engineering, Zhejiang University, Hangzhou 310027, China }
\date{\today}

\begin{abstract}
We propose a graphene-on-grating nanostructure to enable second-order spatial differentiation computation in terahertz (THz) region. The differentiation operation is based on the interference between the direct reflected field and the leakage of two excited surface plasmon polaritons counter-propagating along the graphene sheet. With the spatial coupled-mode theory, we derive out that the requirement for the second-order spatial differentiation is the critical coupling condition. We numerically demonstrate such an analog computation with Gaussian beams. It shows that the spatial bandwidth of the proposed differentiator is large enough such that even when the waist radius of the Gaussian beam is as narrow as ${{w}_{0}}=0.68\lambda $ ($\lambda $ is the free-space wavelength), the accuracy of the differentiator is higher than 95\%. The proposed differentiator is ultra-compact, with a thickness less than $0.1\lambda $, and useful for real-time imaging applications in THz security detections.
\end{abstract}
\maketitle

Terahertz electromagnetic wave exhibits several unique properties. It can penetrate barriers such as clothing and packing materials and is used for spectral characterization of resonances in meV range, such as phonon rotational and vibrational modes in solid substances. Importantly, nonionizing radiation promises THz waves to be harmless \cite{Hu1995,Jepsen11,Ferguson2002,Karpowicz2005,Shen2005}. Therefore, THz wave is suitable for a wide range of imaging applications, especially for contact-free security scanning \cite{Hu1995,Jepsen11,Ferguson2002,Karpowicz2005,Shen2005,Kawase2003,tonouchi2007cet}. However, in THz real-time scanning applications, the high-throughput image processing demands time-consuming computation, which represents a key challenge in practice \cite{Jepsen11}. In the past decade, an impressive range of photonic devices performing analog computing have been proposed to improve information processing speed several-order higher than their electronic counterparts \cite{Kulishov2005,Berger2007,Park2007,bykov2011temporal,wu2014compact,Preciado2008,Ferrera2010}. Especially, optical spatial differentiators are of great interests in image applications, which are capable of detecting edges in an entire image with a single shot \cite{doskolovich2014spatial,Bykov2014,Golovastikov2015,Silva2014performing,AbdollahRamezani2015,Youssefi16,Chizari2016,HwangDavis16,ZhangWeixuan2016}. Recently, we experimentally demonstrated optical edge detection with a plasmonic differentiator operating in visible region \cite{ZhuTengfeng2017}. Given the ultrafast and high-throughput features of optical analog computation, it can be quite useful to design and realize optical spatial differentiators working in THz region.

In this Letter, we propose a graphene-on-grating nanostructure to realize the second-order spatial differentiation in THz region. We demonstrate that for the normal incidence case, the reflected field corresponds to the second-order spatial differentiation of the incident field. Such an analog computation results from the spatial mode interference between the direct reflected wave and the leakage of two excited surface plasmon polaritons (SPPs) counter-propagating along the graphene sheet. By developing the spatial coupled-mode theory (CMT), we show that the second-order spatial differentiation is realized when the coupling process satisfies the critical coupling condition. The thickness of the proposed device is less than $0.1\lambda $ and such ultra-compact due to the highly confined nature of SPPs on graphene. By numerical simulations, we investigate the performance of the differentiator using Gaussian beams with various waist radius ${{w}_{0}}$. We show that the proposed device has a broad spatial spectral bandwidth, being able to process Gaussian beams as narrow as ${{w}_{0}}=0.68\lambda $, with an accuracy higher than 95\%.

\begin{figure*}[htbp]
\centering
\fbox{\includegraphics[width=\linewidth]{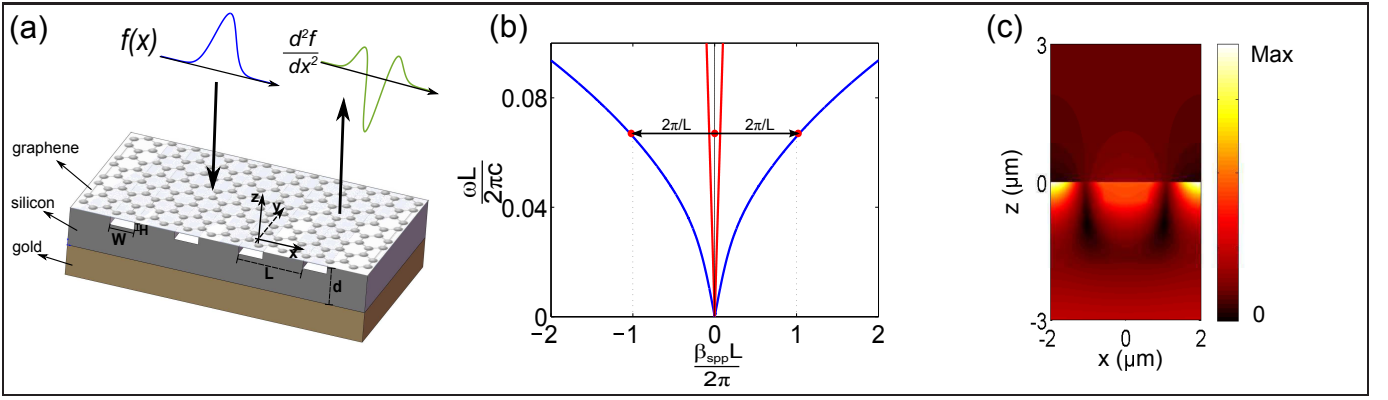}}
\caption{(a) Schematic of the graphene-on-grating nanostructure. Geometric parameters of the structure: $L$=4$\mu m$, $W$=1119nm, $H$=200nm, and $d$=5$\mu m$. (b) Dispersion relation of the SPPs sustained on the air-graphene-Si interface (blue line), and light line cone in Si dielectric (red line). (c) Amplitude distribution of the magnetic field (${{H}_{y}}$) near the interface ($z=0$) for the normal incidence case of TM-polarization where the mangetic field has only $y$-component.}
\label{fig:fig1}
\end{figure*}

Fig.\ref{fig:fig1}(a) schematically shows the graphene-on-grating structure of the proposed differentiator. It can be fabricated by depositing silicon on a thick gold layer, and then patterned and etched as a diffraction grating. A monolayer graphene is grown on copper substrate and coated by PMMA and then transferred onto the Si diffraction grating \cite{LiXuesong2009,ZhuXiaolong2013}. Here the gold layer is assumed thick enough and perfectly conducting in THz frequency region.

Such a monolayer graphene supports SPPs for TM-polarized wave (magnetic field perpendicular to the incident plane), where the collective oscillation of Dirac fermions is in resonance with the electromagnetic field \cite{GarciadeAbajo2014}. The corresponding dispersion diagram of the SPPs on the air-graphene-Si interface is shown in Fig.\ref{fig:fig1}(b). Here the surface conductivity of monolayer graphene is described by the Drude model $\sigma =\frac{{{e}^{2}}}{\pi {{\hbar }^{2}}}\frac{i{{\mu }_{c}}}{\omega +i\gamma }$, where ${{\mu }_{c}}$ and $\gamma $ correspond to the chemical potential and scattering rate, respectively, and $\gamma =\frac{2\pi }{\tau }$ with a finite transport scattering time $\tau $ \cite{GarciadeAbajo2014}. The chemical potential and scattering time of the monolayer graphene are assumed to be ${{\mu }_{c}}=0.6eV$ and $\tau ={{10}^{-11}}s$. The refractive index of the silicon is ${{n}_{Si}}=3.4164$. We note that the dispersion line of graphene SPPs lies far away from the light line cone, i.e. ${{\beta }_{spp}}\gg {{k}_{0}}$, indicating that the excited SPPs are strongly confined to the graphene surface.

To show the physical mechanism of the proposed spatial differentiator, we develop a spatial coupled-mode theory. Here we use the grating coupling method \cite{ZhuXiaolong2013,GaoWeilu2012} to excite the SPPs on graphene, where the period of the grating $L$ is designed to satisfy the phase-matching condition ${{\beta }_{spp}}=\frac{2\pi }{L}$ and ${{\beta }_{spp}}$ is the SPP wavevector. Fig.\ref{fig:fig1}(c) shows the magnetic field at the air-graphene-grating interface for the normal incidence case, and indeed exhibits the strong confinement feature of graphene SPPs. In this case, the incident light simultaneously generates two SPP modes that propagate on the graphene surface along the positive and negative $x$ direction, respectively. The phase matching also enables the two counter-propagating SPP modes to couple with each other. Meanwhile, the propagating SPP modes leak out into air through the phase matching. Therefore, the reflected field distribution results from the interference of three contributions: the direct reflection of the incident wave, and the leakage from two propagating SPP modes. Based on the spatial coupled-mode theory \cite{LouPanZhuRuan16,Ruan2014Spatial,ruan2015spatial}, the spatial-mode coupling and interference process can be described as:
\begin{widetext}
\begin{equation}
\frac{d}{dx}\left( \begin{matrix}
   {{a}_{1}}(x)  \\
   {{a}_{2}}(x)  \\
\end{matrix} \right)=\left[ i\left( \begin{matrix}
   {{\beta }_{spp}} & {{\beta }_{12}}{{e}^{2i\frac{2\pi }{L}x}}  \\
   -\beta _{12}^{\text{*}}{{e}^{-2i\frac{2\pi }{L}x}} & -{{\beta }_{spp}}  \\
\end{matrix} \right)-\left( \begin{matrix}
   {{\alpha }_{l}} & {{\alpha }_{12}}{{e}^{2i\frac{2\pi }{L}x}}  \\
   -\alpha _{12}^{*}{{e}^{-2i\frac{2\pi }{L}x}} & -{{\alpha }_{l}}  \\
\end{matrix} \right)-\left( \begin{matrix}
   {{\alpha }_{0}} & 0  \\
   0 & -{{\alpha }_{0}}  \\
\end{matrix} \right) \right]\left( \begin{matrix}
   {{a}_{1}}(x)  \\
   {{a}_{2}}(x)  \\
\end{matrix} \right)+\left( \begin{matrix}
   \kappa {{e}^{i\frac{2\pi }{L}x}}  \\
   -\kappa {{e}^{-i\frac{2\pi }{L}x}}  \\
\end{matrix} \right){{s}_{+}}(x)
\label{eq:eq1}
\end{equation}
\begin{equation}
{{s}_{-}}(x)={{e}^{i\varphi }}{{s}_{+}}(x)+d{{e}^{-i\frac{2\pi }{L}x}}{{a}_{1}}(x)+d{{e}^{i\frac{2\pi }{L}x}}{{a}_{2}}(x)
\label{eq:eq2}
\end{equation}
\end{widetext}
Here we take the origin of the $x$-coordinate at the middle of the grating slot and the time convention of ${{e}^{-i\omega t}}$, where $\omega $ is the angular frequency of the incident wave. ${{s}_{\pm }}(x)$ and ${{a}_{1,2}}(x)$ are the amplitudes of the incident and reflected magnetic fields and two SPP modes, which are normalized to the $x$-component of Poynting vector and the $x$-direction energy flow respectively \cite{LouPanZhuRuan16}. The phase shift terms ${{exp}(\pm i\frac{2\pi }{L}x)}$ and ${{exp}(\pm 2i\frac{2\pi }{L}x)}$ in Eq.(\ref{eq:eq1},\ref{eq:eq2}) enable the phase-matching between the incident wave and the excited SPP modes propagating in two directions. ${{\alpha }_{l}}$ and ${{\alpha }_{0}}$ represent the loss rate resulting from the leakage radiation of SPPs and the intrinsic loss rate from the material loss, and ${{\beta }_{12}}$ represents the coupling between two excited SPPs. ${{e}^{i\varphi }}$ is the background reflection coefficient without exciting the SPP modes. We note that  the background phase term ${{e}^{i\varphi }}$, the coupling coefficient ${{d}_{0}}$, the leakage rate ${{\alpha }_{l}}$, the cross-coupling terms ${{\beta }_{12}}$ and ${{\alpha }_{12}}$ are not independent of each other, since they are constrained by the energy conservation, mirror-symmetry and time-reversal condition \cite{LouPanZhuRuan16}. As the results of the theory \cite{LouPanZhuRuan16}, we show that both $\beta _{12}$ and $\alpha _{12}$ are real numbers and these parameters are related by
\begin{subequations}
\begin{equation}
{{\beta }_{12}}\text{=}\beta _{12}^{\text{*}}
\end{equation}
\begin{equation}
{{\alpha }_{l}}={{\alpha }_{12}}=\alpha _{12}^{*}=\frac{1}{2}d{{d}^{\text{*}}}
\label{eq:eq3}
\end{equation}
\begin{equation}
\kappa =d
\label{eq:eq4}
\end{equation}
\begin{equation}
{{e}^{i\varphi }}{{d}^{*}}+d=0
\label{eq:eq5}
\end{equation}
\end{subequations}
Eq.(\ref{eq:eq3},\ref{eq:eq5}) further give that $d=\sqrt{2{{\alpha }_{l}}}{{e}^{i(\varphi /2-\pi /2+n\pi )}}$, where $n$ is an integer determined by the choice of the origin point of $x$-axis. Note that the spatial coupled-mode theory takes the approximation of the strong confinement condition, ${{\alpha }_{l}}+{{\alpha }_{0}}\ll {{\beta }_{spp}}$ \cite{Ruan2014Spatial,haus1984waves,fan2003temporal}.

Based on Eq.(\ref{eq:eq1},\ref{eq:eq2}), we obtain the spatial spectral transfer function for the graphene-on-grating structure. We expand the incident and the reflected field into a series of plane waves as ${{s}_{\pm }}(x)=\int{{{\widetilde{s}}_{\pm }}({{k}_{x}}){{e}^{ik{}_{x}x}}d{{k}_{x}}}$, where ${{\widetilde{s}}_{\pm }}({{k}_{x}})$ is the amplitude of each plane wave and ${{k}_{x}}$ represents the $x$-component of the wavevector. By transforming Eq.(\ref{eq:eq1},\ref{eq:eq2}) into the Fourier domain, the spatial spectral transfer function is obtained as
\begin{equation}
H({{k}_{x}})\equiv \frac{{{\widetilde{s}}_{-}}({{k}_{x}})}{{{\widetilde{s}}_{+}}({{k}_{x}})}={{e}^{i\varphi }}\frac{k_{x}^{2}+(-2{{\alpha }_{l}}+{{\alpha }_{0}}){{\alpha }_{0}}-2i{{\alpha }_{l}}{{\beta }_{12}}+\beta _{12}^{2}}{k_{x}^{2}+(2{{\alpha }_{l}}+{{\alpha }_{0}}){{\alpha }_{0}}+2i{{\alpha }_{l}}{{\beta }_{12}}+\beta _{12}^{2}}
\label{eq:eq6}
\end{equation}
We note that in the lossless case ${{\alpha }_{0}}=0$, $\left| H({{k}_{x}}) \right|=1$, which is consistent with the energy conservation condition. Especially, when the critical coupling condition ${{\alpha }_{0}}=2{{\alpha }_{l}}$ is satisfied, and ${{\beta }_{12}}$ is small enough to be approximated to zero, the transfer function can be approximated in $\left| {{k}_{x}} \right|\ll 2{{\alpha }_{l}}$ as
\begin{equation}
H({{k}_{x}})\approx \frac{{{e}^{i\varphi }}}{8\alpha _{0}^{2}}k_{x}^{2}
\label{eq:eq7}
\end{equation}
Eq.(\ref{eq:eq7}) is the spatial frequency domain transfer function for a spatial second-order differentiation, which has quadratic dependence of ${{k}_{x}}$ at about ${{k}_{x}}=0$. Correspondingly, in the spatial domain, the reflected field profile is proportional to the spatial second-order differentiation as
\begin{equation}
{{s}_{-}}=\frac{{{e}^{i\varphi }}}{8\alpha _{0}^{2}}\frac{{{d}^{2}}{{s}_{+}}}{d{{x}^{2}}}
\label{eq:eq8}
\end{equation}

\begin{figure}[htbp]
\centering
\fbox{\includegraphics[width=\linewidth]{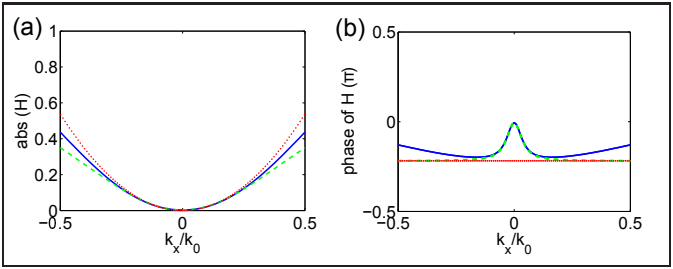}}
\caption{(a) Amplitude and (b) phase spectra of transfer function. The numerical simulations (blue solid lines), the CMT fitting results (green dash lines), and the ideal differentiator case for Eq.(\ref{eq:eq7}) (red dotted lines).}
\label{fig:fig2}
\end{figure}

To realize the spatial second-order differentiation we design the depth of the grating slot to satisfy the critical coupling condition ${{\alpha }_{0}}=2{{\alpha }_{l}}$. Here we consider that the incident wave is at 5.368 THz. The intrinsic material loss of the SPP ${{\alpha }_{0}}$ is mainly  determined by the graphene conductivity and thus is insensitive to the slot size. On the other hand, the leakage loss ${{\alpha }_{l}}$ of the SPP monotonically increases as the slot width and depth increase. Thus, the critical coupling condition can be realized by appropriately designing the slot width and depth of the grating structure. Guided by these criterions, we design a graphene-on-grating structure where the Si dielectric layer has a slot width $W=1119$nm, slot depth $H=200$nm, thickness $d=5$$\mu m$, and period $L=4$$\mu m$.

In order to demonstrate the spatial second-order differentiation, we first compare the spatial spectral transfer function of the proposed graphene-on-grating structure with the the one of an ideal spatial second-order differentiator (Fig.\ref{fig:fig2}). We numerically calculate the transfer function by the finite element method using the commercial software COMSOL, and the results are shown as the blue solid lines in Fig.\ref{fig:fig2}. To validate our CMT theory, we fit the calculated transfer function with Eq.(\ref{eq:eq6}) (the green dashed lines in Fig.\ref{fig:fig2}). It shows that the amplitude of the transfer function can be well fitted with the parameters ${{\alpha }_{l}}=0.241{{k}_{0}}$, ${{\alpha }_{0}}=2.01{{\alpha }_{l}}$ and  ${{\beta }_{12}}=-0.008{{\alpha }_{l}}$ in the range $\left| {{k}_{x}} \right|<0.25{{k}_{0}}$, where $k_0$ is the free space wavevector. We note that the phase of the transfer function exhibits a peak at the vicinity of ${{k}_{x}}=0$. It is well fitted by the CMT under the consideration of the cross-coupling of two SPP modes described by ${{\beta }_{12}}$. We also plot the ideal transfer function of a second-order differentiator by Eq.(\ref{eq:eq7}) with the fitting parameters ${{\alpha }_{0}}=2.01{{\alpha }_{l}}$.  In comparison with the numerical results, it confirms that the transfer function of the proposed graphene plasmonic structure exhibits a quadratic dependence of ${{k}_{x}}$ near ${{k}_{x}}=0$. Moreover, as our numerical simulations show below, the phase difference between the real and the ideal cases has a rather minor impact on the differentiation accuracy.

\begin{figure}[htbp]
\centering
\fbox{\includegraphics[width=\linewidth]{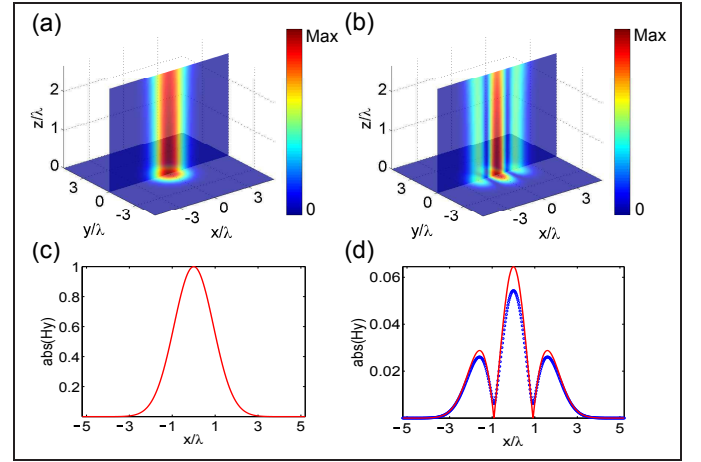}}
\caption{Amplitude distribution of the magnetic field for (a) incident 3D TM-polarized Gaussian beam ${{H}_{y}}={{e}^{-({{x}^{2}}+{{y}^{2}})/w_{0}^{2}}}$ and (b) the simulated reflected beam. The incident Gaussian beam has a waist radius ${{w}_{0}}=1.3\lambda $  and focuses on the interface $z=0$. (c) The amplitude of the incident field along the $x$-axis at the interface $z=0$. (d) The amplitude of the reflected field along the $x$-axis at the interface $z=0$ (blue circles), and ideal differentiation analytically computed by Eq.(\ref{eq:eq8})(red line).}
\label{fig:fig3}
\end{figure}

We now illustrate the spatial second-order differentiation with a Gaussian beam illumination. The incident beam has a TM-polarized magnetic field profile ${{H}_{y}}={{e}^{-({{x}^{2}}+{{y}^{2}})/w_{0}^{2}}}$ with a waist radius ${{w}_{0}}=1.3\lambda $ and focuses on the grating normally. Fig.\ref{fig:fig3}(a) and \ref{fig:fig3}(b) show the $y$ component of magnetic field (${{H}_{y}}$) for the incident and the reflected beams. Here we calculate the reflected field using the three-dimension full vector Fourier optics method where the reflection coefficients for each plane wave are calculated with COMSOL. Fig.\ref{fig:fig3}(b) exhibits three peaks in the reflected beam, which are the obvious features of a second-order spatial differentiation of the incident Gaussian beam on $x$-direction. For a more explicit presentation, we specifically extract the reflected ${{H}_{y}}$ field amplitude along $x$-axis at the interface $z=0$, as plotted in Fig.\ref{fig:fig3}(d). The ideal second-order spatial differentiation result analytically computed using Eq.(\ref{eq:eq8}) is also plotted for comparison. The simulated reflected field amplitude agrees well with the analytical differentiation result. The accuracy of the differentiation is $99.8\%$, described by the Pearson correlation coefficient between the simulated and analytical computed reflected field amplitudes along $x$-axis at the interface.
\begin{figure}[htbp]
\centering
\fbox{\includegraphics[width=\linewidth]{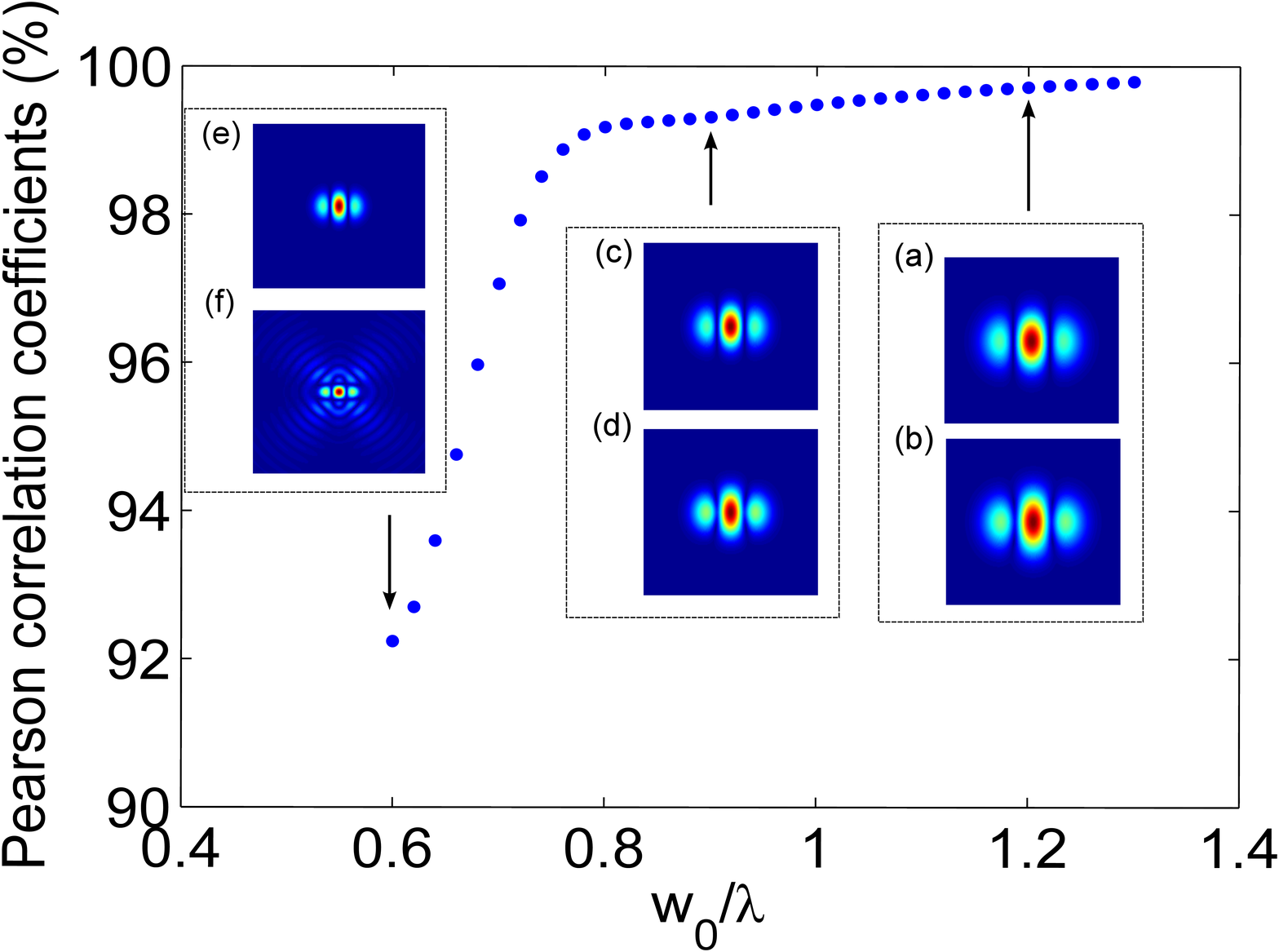}}
\caption{Pearson correlation coefficients between the simulated reflected field amplitudes and analytical differentiation results with respect to ${{w}_{0}}/\lambda $. Inset: The numerically simulated amplitude of the reflected field ${{H}_{y}}$ at the interface $z=0$ (b,d,f), and the corresponding ideal ones for the second-order spatial differentiation (a,c,e), for ${{w}_{0}}/\lambda =$ 1.2, 0.9, and 0.6 respectively.}
\label{fig:fig4}
\end{figure}

Eq.(\ref{eq:eq6}) shows that the spatial bandwidth of the differentiator is limited by the leakage loss rate ${{\alpha }_{l}}$. Here in comparison with the free space wavevector $k_0$ the intrinsic loss rate ${{\alpha }_{0}}$ is $0.241{{k}_{0}}$, which indicates that the differentiator has a broad operation bandwidth such that it is capable of resolving the change of the incident field when the beam size is small. To investigate the spatial resolution of our plasmonic differentiator, we gradually reduce the beam size and simulate the field transformation during the reflection. Fig.\ref{fig:fig4} shows  the Pearson correlation coefficients with respect to ${{w}_{0}}/\lambda $. The device implements second-order spatial differentiation with the Pearson correlation coefficients over $99\%$ for incident Gaussian beams with ${{w}_{0}}>\lambda$. When Gaussian beam waist radius ${{w}_{0}}$ becomes narrower, the differentiation result degrades. The insets of Fig.\ref{fig:fig4} (b), (d) and (f) show the numerically simulated reflected field patterns at the interface $z=0$, for ${{w}_{0}}/\lambda =$ 1.2, 0.9 and 0.6 respectively. Correspondingly, Fig.\ref{fig:fig4} (a), (c) and (e) show the ideal ones for the second-order spatial differentiation. We note that even when the Gaussian beam size is as narrow as ${{w}_{0}}/\lambda =0.68$ , nearly the diffraction limit, the differentiator still works effectively, with the Pearson correlation coefficient over $95\%$. This feature shows that the plasmonic differentiator has a very high spatial resolution, which is useful for applications in THz image sharpening and edge detections.

In summary, we have demonstrated that the proposed graphene-on-grating nanostructure can perform second-order spatial differentiation on THz waves at normal incidence. The desired differentiation is realized when two counter-propagating SPPs on graphene surface are excited and the critical coupling condition ${{\alpha }_{0}}=2{{\alpha }_{l}}$ is satisfied. The device is ultra-compact and has a broad spatial operation bandwidth, which promises it the ability to process ultra-narrow optical signals, for example, Gaussian beam as narrow as ${{w}_{0}}=0.68\lambda $. Such a miniaturized and broadband photonic second-order spatial differentiator could be useful for high-resolution all-optical signal processing and imaging applications in THz region.

The authors acknowledge the financial support by Fundamental Research Funds for the Central Universities (2014QNA3007), and the National Natural Science Foundation of China (NSFC 61675179).


\begin{thebibliography}{33}%
\makeatletter
\providecommand \@ifxundefined [1]{%
 \@ifx{#1\undefined}
}%
\providecommand \@ifnum [1]{%
 \ifnum #1\expandafter \@firstoftwo
 \else \expandafter \@secondoftwo
 \fi
}%
\providecommand \@ifx [1]{%
 \ifx #1\expandafter \@firstoftwo
 \else \expandafter \@secondoftwo
 \fi
}%
\providecommand \natexlab [1]{#1}%
\providecommand \enquote  [1]{``#1''}%
\providecommand \bibnamefont  [1]{#1}%
\providecommand \bibfnamefont [1]{#1}%
\providecommand \citenamefont [1]{#1}%
\providecommand \href@noop [0]{\@secondoftwo}%
\providecommand \href [0]{\begingroup \@sanitize@url \@href}%
\providecommand \@href[1]{\@@startlink{#1}\@@href}%
\providecommand \@@href[1]{\endgroup#1\@@endlink}%
\providecommand \@sanitize@url [0]{\catcode `\\12\catcode `\$12\catcode
  `\&12\catcode `\#12\catcode `\^12\catcode `\_12\catcode `\%12\relax}%
\providecommand \@@startlink[1]{}%
\providecommand \@@endlink[0]{}%
\providecommand \url  [0]{\begingroup\@sanitize@url \@url }%
\providecommand \@url [1]{\endgroup\@href {#1}{\urlprefix }}%
\providecommand \urlprefix  [0]{URL }%
\providecommand \Eprint [0]{\href }%
\providecommand \doibase [0]{http://dx.doi.org/}%
\providecommand \selectlanguage [0]{\@gobble}%
\providecommand \bibinfo  [0]{\@secondoftwo}%
\providecommand \bibfield  [0]{\@secondoftwo}%
\providecommand \translation [1]{[#1]}%
\providecommand \BibitemOpen [0]{}%
\providecommand \bibitemStop [0]{}%
\providecommand \bibitemNoStop [0]{.\EOS\space}%
\providecommand \EOS [0]{\spacefactor3000\relax}%
\providecommand \BibitemShut  [1]{\csname bibitem#1\endcsname}%
\let\auto@bib@innerbib\@empty
\bibitem [{\citenamefont {Hu}\ and\ \citenamefont {Nuss}(1995)}]{Hu1995}%
  \BibitemOpen
  \bibfield  {author} {\bibinfo {author} {\bibfnamefont {B.~B.}\ \bibnamefont
  {Hu}}\ and\ \bibinfo {author} {\bibfnamefont {M.~C.}\ \bibnamefont {Nuss}},\
  }\href@noop {} {\bibfield  {journal} {\bibinfo  {journal} {Optics Letters}\
  }\textbf {\bibinfo {volume} {20}},\ \bibinfo {pages} {1716} (\bibinfo {year}
  {1995})}\BibitemShut {NoStop}%
\bibitem [{\citenamefont {Jepsen}\ \emph {et~al.}(2011)\citenamefont {Jepsen},
  \citenamefont {Cooke},\ and\ \citenamefont {Koch}}]{Jepsen11}%
  \BibitemOpen
  \bibfield  {author} {\bibinfo {author} {\bibfnamefont {P.~U.}\ \bibnamefont
  {Jepsen}}, \bibinfo {author} {\bibfnamefont {D.~G.}\ \bibnamefont {Cooke}}, \
  and\ \bibinfo {author} {\bibfnamefont {M.}~\bibnamefont {Koch}},\ }\href@noop
  {} {\bibfield  {journal} {\bibinfo  {journal} {Laser \& Photonics Reviews}\
  }\textbf {\bibinfo {volume} {5}},\ \bibinfo {pages} {124} (\bibinfo {year}
  {2011})}\BibitemShut {NoStop}%
\bibitem [{\citenamefont {Ferguson}\ and\ \citenamefont
  {Zhang}(2002)}]{Ferguson2002}%
  \BibitemOpen
  \bibfield  {author} {\bibinfo {author} {\bibfnamefont {B.}~\bibnamefont
  {Ferguson}}\ and\ \bibinfo {author} {\bibfnamefont {X.~C.}\ \bibnamefont
  {Zhang}},\ }\href@noop {} {\bibfield  {journal} {\bibinfo  {journal} {Nature
  materials}\ }\textbf {\bibinfo {volume} {1}},\ \bibinfo {pages} {26}
  (\bibinfo {year} {2002})}\BibitemShut {NoStop}%
\bibitem [{\citenamefont {Karpowicz}\ \emph {et~al.}(2005)\citenamefont
  {Karpowicz}, \citenamefont {Zhong}, \citenamefont {Zhang}, \citenamefont
  {Lin}, \citenamefont {Hwang}, \citenamefont {Xu},\ and\ \citenamefont
  {Zhang}}]{Karpowicz2005}%
  \BibitemOpen
  \bibfield  {author} {\bibinfo {author} {\bibfnamefont {N.}~\bibnamefont
  {Karpowicz}}, \bibinfo {author} {\bibfnamefont {H.}~\bibnamefont {Zhong}},
  \bibinfo {author} {\bibfnamefont {C.~L.}\ \bibnamefont {Zhang}}, \bibinfo
  {author} {\bibfnamefont {K.~I.}\ \bibnamefont {Lin}}, \bibinfo {author}
  {\bibfnamefont {J.~S.}\ \bibnamefont {Hwang}}, \bibinfo {author}
  {\bibfnamefont {J.~Z.}\ \bibnamefont {Xu}}, \ and\ \bibinfo {author}
  {\bibfnamefont {X.~C.}\ \bibnamefont {Zhang}},\ }\href@noop {} {\bibfield
  {journal} {\bibinfo  {journal} {Applied Physics Letters}\ }\textbf {\bibinfo
  {volume} {86}},\ \bibinfo {pages} {054105} (\bibinfo {year}
  {2005})}\BibitemShut {NoStop}%
\bibitem [{\citenamefont {Shen}\ \emph {et~al.}(2005)\citenamefont {Shen},
  \citenamefont {Lo}, \citenamefont {Taday}, \citenamefont {Cole},
  \citenamefont {Tribe},\ and\ \citenamefont {Kemp}}]{Shen2005}%
  \BibitemOpen
  \bibfield  {author} {\bibinfo {author} {\bibfnamefont {Y.~C.}\ \bibnamefont
  {Shen}}, \bibinfo {author} {\bibfnamefont {T.}~\bibnamefont {Lo}}, \bibinfo
  {author} {\bibfnamefont {P.~F.}\ \bibnamefont {Taday}}, \bibinfo {author}
  {\bibfnamefont {B.~E.}\ \bibnamefont {Cole}}, \bibinfo {author}
  {\bibfnamefont {W.~R.}\ \bibnamefont {Tribe}}, \ and\ \bibinfo {author}
  {\bibfnamefont {M.~C.}\ \bibnamefont {Kemp}},\ }\href@noop {} {\bibfield
  {journal} {\bibinfo  {journal} {Applied Physics Letters}\ }\textbf {\bibinfo
  {volume} {86}},\ \bibinfo {pages} {241116} (\bibinfo {year}
  {2005})}\BibitemShut {NoStop}%
\bibitem [{\citenamefont {Kawase}\ \emph {et~al.}(2003)\citenamefont {Kawase},
  \citenamefont {Ogawa}, \citenamefont {Watanabe},\ and\ \citenamefont
  {Inoue}}]{Kawase2003}%
  \BibitemOpen
  \bibfield  {author} {\bibinfo {author} {\bibfnamefont {K.}~\bibnamefont
  {Kawase}}, \bibinfo {author} {\bibfnamefont {Y.}~\bibnamefont {Ogawa}},
  \bibinfo {author} {\bibfnamefont {Y.}~\bibnamefont {Watanabe}}, \ and\
  \bibinfo {author} {\bibfnamefont {H.}~\bibnamefont {Inoue}},\ }\href@noop {}
  {\bibfield  {journal} {\bibinfo  {journal} {Optics Express}\ }\textbf
  {\bibinfo {volume} {11}},\ \bibinfo {pages} {2549} (\bibinfo {year}
  {2003})}\BibitemShut {NoStop}%
\bibitem [{\citenamefont {Tonouchi}(2007)}]{tonouchi2007cet}%
  \BibitemOpen
  \bibfield  {author} {\bibinfo {author} {\bibfnamefont {M.}~\bibnamefont
  {Tonouchi}},\ }\href@noop {} {\bibfield  {journal} {\bibinfo  {journal}
  {Nature Photonics}\ }\textbf {\bibinfo {volume} {1}},\ \bibinfo {pages} {97}
  (\bibinfo {year} {2007})}\BibitemShut {NoStop}%
\bibitem [{\citenamefont {Kulishov}\ and\ \citenamefont
  {Azana}(2005)}]{Kulishov2005}%
  \BibitemOpen
  \bibfield  {author} {\bibinfo {author} {\bibfnamefont {M.}~\bibnamefont
  {Kulishov}}\ and\ \bibinfo {author} {\bibfnamefont {J.}~\bibnamefont
  {Azana}},\ }\href@noop {} {\bibfield  {journal} {\bibinfo  {journal} {Optics
  Letters}\ }\textbf {\bibinfo {volume} {30}},\ \bibinfo {pages} {2700}
  (\bibinfo {year} {2005})}\BibitemShut {NoStop}%
\bibitem [{\citenamefont {Berger}\ \emph {et~al.}(2007)\citenamefont {Berger},
  \citenamefont {Levit}, \citenamefont {Fischer}, \citenamefont {Kulishov},
  \citenamefont {Plant},\ and\ \citenamefont {Azana}}]{Berger2007}%
  \BibitemOpen
  \bibfield  {author} {\bibinfo {author} {\bibfnamefont {N.~K.}\ \bibnamefont
  {Berger}}, \bibinfo {author} {\bibfnamefont {B.}~\bibnamefont {Levit}},
  \bibinfo {author} {\bibfnamefont {B.}~\bibnamefont {Fischer}}, \bibinfo
  {author} {\bibfnamefont {M.}~\bibnamefont {Kulishov}}, \bibinfo {author}
  {\bibfnamefont {D.~V.}\ \bibnamefont {Plant}}, \ and\ \bibinfo {author}
  {\bibfnamefont {J.}~\bibnamefont {Azana}},\ }\href@noop {} {\bibfield
  {journal} {\bibinfo  {journal} {Optics Express}\ }\textbf {\bibinfo {volume}
  {15}},\ \bibinfo {pages} {371} (\bibinfo {year} {2007})}\BibitemShut
  {NoStop}%
\bibitem [{\citenamefont {Park}\ \emph {et~al.}(2007)\citenamefont {Park},
  \citenamefont {Azana},\ and\ \citenamefont {Slavik}}]{Park2007}%
  \BibitemOpen
  \bibfield  {author} {\bibinfo {author} {\bibfnamefont {Y.}~\bibnamefont
  {Park}}, \bibinfo {author} {\bibfnamefont {J.}~\bibnamefont {Azana}}, \ and\
  \bibinfo {author} {\bibfnamefont {R.}~\bibnamefont {Slavik}},\ }\href@noop {}
  {\bibfield  {journal} {\bibinfo  {journal} {Optics Letters}\ }\textbf
  {\bibinfo {volume} {32}},\ \bibinfo {pages} {710} (\bibinfo {year}
  {2007})}\BibitemShut {NoStop}%
\bibitem [{\citenamefont {Bykov}\ \emph {et~al.}(2011)\citenamefont {Bykov},
  \citenamefont {Doskolovich},\ and\ \citenamefont
  {Soifer}}]{bykov2011temporal}%
  \BibitemOpen
  \bibfield  {author} {\bibinfo {author} {\bibfnamefont {D.~A.}\ \bibnamefont
  {Bykov}}, \bibinfo {author} {\bibfnamefont {L.~L.}\ \bibnamefont
  {Doskolovich}}, \ and\ \bibinfo {author} {\bibfnamefont {V.~A.}\ \bibnamefont
  {Soifer}},\ }\href@noop {} {\bibfield  {journal} {\bibinfo  {journal} {Optics
  Letters}\ }\textbf {\bibinfo {volume} {36}},\ \bibinfo {pages} {3509}
  (\bibinfo {year} {2011})}\BibitemShut {NoStop}%
\bibitem [{\citenamefont {Wu}\ \emph {et~al.}(2014)\citenamefont {Wu},
  \citenamefont {Cao}, \citenamefont {Hu}, \citenamefont {Jiang}, \citenamefont
  {Pan}, \citenamefont {Yang}, \citenamefont {Qiu}, \citenamefont {Tremblay},\
  and\ \citenamefont {Su}}]{wu2014compact}%
  \BibitemOpen
  \bibfield  {author} {\bibinfo {author} {\bibfnamefont {J.}~\bibnamefont
  {Wu}}, \bibinfo {author} {\bibfnamefont {P.}~\bibnamefont {Cao}}, \bibinfo
  {author} {\bibfnamefont {X.}~\bibnamefont {Hu}}, \bibinfo {author}
  {\bibfnamefont {X.}~\bibnamefont {Jiang}}, \bibinfo {author} {\bibfnamefont
  {T.}~\bibnamefont {Pan}}, \bibinfo {author} {\bibfnamefont {Y.}~\bibnamefont
  {Yang}}, \bibinfo {author} {\bibfnamefont {C.}~\bibnamefont {Qiu}}, \bibinfo
  {author} {\bibfnamefont {C.}~\bibnamefont {Tremblay}}, \ and\ \bibinfo
  {author} {\bibfnamefont {Y.}~\bibnamefont {Su}},\ }\href@noop {} {\bibfield
  {journal} {\bibinfo  {journal} {Optics Express}\ }\textbf {\bibinfo {volume}
  {22}},\ \bibinfo {pages} {26254} (\bibinfo {year} {2014})}\BibitemShut
  {NoStop}%
\bibitem [{\citenamefont {Preciado}\ and\ \citenamefont
  {Muriel}(2008)}]{Preciado2008}%
  \BibitemOpen
  \bibfield  {author} {\bibinfo {author} {\bibfnamefont {M.~A.}\ \bibnamefont
  {Preciado}}\ and\ \bibinfo {author} {\bibfnamefont {M.~A.}\ \bibnamefont
  {Muriel}},\ }\href@noop {} {\bibfield  {journal} {\bibinfo  {journal} {Optics
  Letters}\ }\textbf {\bibinfo {volume} {33}},\ \bibinfo {pages} {1348}
  (\bibinfo {year} {2008})}\BibitemShut {NoStop}%
\bibitem [{\citenamefont {Ferrera}\ \emph {et~al.}(2010)\citenamefont
  {Ferrera}, \citenamefont {Park}, \citenamefont {Razzari}, \citenamefont
  {Little}, \citenamefont {Chu}, \citenamefont {Morandotti}, \citenamefont
  {Moss},\ and\ \citenamefont {Azana}}]{Ferrera2010}%
  \BibitemOpen
  \bibfield  {author} {\bibinfo {author} {\bibfnamefont {M.}~\bibnamefont
  {Ferrera}}, \bibinfo {author} {\bibfnamefont {Y.}~\bibnamefont {Park}},
  \bibinfo {author} {\bibfnamefont {L.}~\bibnamefont {Razzari}}, \bibinfo
  {author} {\bibfnamefont {B.~E.}\ \bibnamefont {Little}}, \bibinfo {author}
  {\bibfnamefont {S.~T.}\ \bibnamefont {Chu}}, \bibinfo {author} {\bibfnamefont
  {R.}~\bibnamefont {Morandotti}}, \bibinfo {author} {\bibfnamefont {D.~J.}\
  \bibnamefont {Moss}}, \ and\ \bibinfo {author} {\bibfnamefont
  {J.}~\bibnamefont {Azana}},\ }\href@noop {} {\bibfield  {journal} {\bibinfo
  {journal} {Nature Communications}\ }\textbf {\bibinfo {volume} {1}},\
  \bibinfo {pages} {29} (\bibinfo {year} {2010})}\BibitemShut {NoStop}%
\bibitem [{\citenamefont {Doskolovich}\ \emph {et~al.}(2014)\citenamefont
  {Doskolovich}, \citenamefont {Bykov}, \citenamefont {Bezus},\ and\
  \citenamefont {Soifer}}]{doskolovich2014spatial}%
  \BibitemOpen
  \bibfield  {author} {\bibinfo {author} {\bibfnamefont {L.~L.}\ \bibnamefont
  {Doskolovich}}, \bibinfo {author} {\bibfnamefont {D.~A.}\ \bibnamefont
  {Bykov}}, \bibinfo {author} {\bibfnamefont {E.~A.}\ \bibnamefont {Bezus}}, \
  and\ \bibinfo {author} {\bibfnamefont {V.~A.}\ \bibnamefont {Soifer}},\
  }\href@noop {} {\bibfield  {journal} {\bibinfo  {journal} {Optics Letters}\
  }\textbf {\bibinfo {volume} {39}},\ \bibinfo {pages} {1278} (\bibinfo {year}
  {2014})}\BibitemShut {NoStop}%
\bibitem [{\citenamefont {Bykov}\ \emph {et~al.}(2014)\citenamefont {Bykov},
  \citenamefont {Doskolovich}, \citenamefont {Bezus},\ and\ \citenamefont
  {Soifer}}]{Bykov2014}%
  \BibitemOpen
  \bibfield  {author} {\bibinfo {author} {\bibfnamefont {D.~A.}\ \bibnamefont
  {Bykov}}, \bibinfo {author} {\bibfnamefont {L.~L.}\ \bibnamefont
  {Doskolovich}}, \bibinfo {author} {\bibfnamefont {E.~A.}\ \bibnamefont
  {Bezus}}, \ and\ \bibinfo {author} {\bibfnamefont {V.~A.}\ \bibnamefont
  {Soifer}},\ }\href@noop {} {\bibfield  {journal} {\bibinfo  {journal} {Optics
  Express}\ }\textbf {\bibinfo {volume} {22}},\ \bibinfo {pages} {25084}
  (\bibinfo {year} {2014})}\BibitemShut {NoStop}%
\bibitem [{\citenamefont {Golovastikov}\ \emph {et~al.}(2015)\citenamefont
  {Golovastikov}, \citenamefont {Bykov},\ and\ \citenamefont
  {Doskolovich}}]{Golovastikov2015}%
  \BibitemOpen
  \bibfield  {author} {\bibinfo {author} {\bibfnamefont {N.~V.}\ \bibnamefont
  {Golovastikov}}, \bibinfo {author} {\bibfnamefont {D.~A.}\ \bibnamefont
  {Bykov}}, \ and\ \bibinfo {author} {\bibfnamefont {L.~L.}\ \bibnamefont
  {Doskolovich}},\ }\href@noop {} {\bibfield  {journal} {\bibinfo  {journal}
  {Optics Letters}\ }\textbf {\bibinfo {volume} {40}},\ \bibinfo {pages} {3492}
  (\bibinfo {year} {2015})}\BibitemShut {NoStop}%
\bibitem [{\citenamefont {Silva}\ \emph {et~al.}(2014)\citenamefont {Silva},
  \citenamefont {Monticone}, \citenamefont {Castaldi}, \citenamefont {Galdi},
  \citenamefont {Al{\`u}},\ and\ \citenamefont
  {Engheta}}]{Silva2014performing}%
  \BibitemOpen
  \bibfield  {author} {\bibinfo {author} {\bibfnamefont {A.}~\bibnamefont
  {Silva}}, \bibinfo {author} {\bibfnamefont {F.}~\bibnamefont {Monticone}},
  \bibinfo {author} {\bibfnamefont {G.}~\bibnamefont {Castaldi}}, \bibinfo
  {author} {\bibfnamefont {V.}~\bibnamefont {Galdi}}, \bibinfo {author}
  {\bibfnamefont {A.}~\bibnamefont {Al{\`u}}}, \ and\ \bibinfo {author}
  {\bibfnamefont {N.}~\bibnamefont {Engheta}},\ }\href@noop {} {\bibfield
  {journal} {\bibinfo  {journal} {Science}\ }\textbf {\bibinfo {volume}
  {343}},\ \bibinfo {pages} {160} (\bibinfo {year} {2014})}\BibitemShut
  {NoStop}%
\bibitem [{\citenamefont {AbdollahRamezani}\ \emph {et~al.}(2015)\citenamefont
  {AbdollahRamezani}, \citenamefont {Arik}, \citenamefont {Khavasi},\ and\
  \citenamefont {Kavehvash}}]{AbdollahRamezani2015}%
  \BibitemOpen
  \bibfield  {author} {\bibinfo {author} {\bibfnamefont {S.}~\bibnamefont
  {AbdollahRamezani}}, \bibinfo {author} {\bibfnamefont {K.}~\bibnamefont
  {Arik}}, \bibinfo {author} {\bibfnamefont {A.}~\bibnamefont {Khavasi}}, \
  and\ \bibinfo {author} {\bibfnamefont {Z.}~\bibnamefont {Kavehvash}},\
  }\href@noop {} {\bibfield  {journal} {\bibinfo  {journal} {Optics Letters}\
  }\textbf {\bibinfo {volume} {40}},\ \bibinfo {pages} {5239} (\bibinfo {year}
  {2015})}\BibitemShut {NoStop}%
\bibitem [{\citenamefont {Youssefi}\ \emph {et~al.}(2016)\citenamefont
  {Youssefi}, \citenamefont {Zangeneh-Nejad}, \citenamefont
  {Abdollahramezani},\ and\ \citenamefont {Khavasi}}]{Youssefi16}%
  \BibitemOpen
  \bibfield  {author} {\bibinfo {author} {\bibfnamefont {A.}~\bibnamefont
  {Youssefi}}, \bibinfo {author} {\bibfnamefont {F.}~\bibnamefont
  {Zangeneh-Nejad}}, \bibinfo {author} {\bibfnamefont {S.}~\bibnamefont
  {Abdollahramezani}}, \ and\ \bibinfo {author} {\bibfnamefont
  {A.}~\bibnamefont {Khavasi}},\ }\href@noop {} {\bibfield  {journal} {\bibinfo
   {journal} {Optics Letters}\ }\textbf {\bibinfo {volume} {41}},\ \bibinfo
  {pages} {3467} (\bibinfo {year} {2016})}\BibitemShut {NoStop}%
\bibitem [{\citenamefont {Chizari}\ \emph {et~al.}(2016)\citenamefont
  {Chizari}, \citenamefont {Abdollahramezani}, \citenamefont {Jamali},\ and\
  \citenamefont {Salehi}}]{Chizari2016}%
  \BibitemOpen
  \bibfield  {author} {\bibinfo {author} {\bibfnamefont {A.}~\bibnamefont
  {Chizari}}, \bibinfo {author} {\bibfnamefont {S.}~\bibnamefont
  {Abdollahramezani}}, \bibinfo {author} {\bibfnamefont {M.~V.}\ \bibnamefont
  {Jamali}}, \ and\ \bibinfo {author} {\bibfnamefont {J.~A.}\ \bibnamefont
  {Salehi}},\ }\href@noop {} {\bibfield  {journal} {\bibinfo  {journal} {Optics
  Letters}\ }\textbf {\bibinfo {volume} {41}},\ \bibinfo {pages} {3451}
  (\bibinfo {year} {2016})}\BibitemShut {NoStop}%
\bibitem [{\citenamefont {Hwang}\ and\ \citenamefont
  {Davis}(2016)}]{HwangDavis16}%
  \BibitemOpen
  \bibfield  {author} {\bibinfo {author} {\bibfnamefont {Y.}~\bibnamefont
  {Hwang}}\ and\ \bibinfo {author} {\bibfnamefont {T.~J.}\ \bibnamefont
  {Davis}},\ }\href@noop {} {\bibfield  {journal} {\bibinfo  {journal} {Applied
  Physics Letters}\ }\textbf {\bibinfo {volume} {109}},\ \bibinfo {pages}
  {181101} (\bibinfo {year} {2016})}\BibitemShut {NoStop}%
\bibitem [{\citenamefont {Zhang}\ \emph {et~al.}(2016)\citenamefont {Zhang},
  \citenamefont {Qu},\ and\ \citenamefont {Zhang}}]{ZhangWeixuan2016}%
  \BibitemOpen
  \bibfield  {author} {\bibinfo {author} {\bibfnamefont {W.}~\bibnamefont
  {Zhang}}, \bibinfo {author} {\bibfnamefont {C.}~\bibnamefont {Qu}}, \ and\
  \bibinfo {author} {\bibfnamefont {X.}~\bibnamefont {Zhang}},\ }\href@noop {}
  {\bibfield  {journal} {\bibinfo  {journal} {Journal of Optics}\ }\textbf
  {\bibinfo {volume} {18}},\ \bibinfo {pages} {075102} (\bibinfo {year}
  {2016})}\BibitemShut {NoStop}%
\bibitem [{\citenamefont {Zhu}\ \emph {et~al.}(2017)\citenamefont {Zhu},
  \citenamefont {Zhou}, \citenamefont {Lou}, \citenamefont {Ye}, \citenamefont
  {Qiu}, \citenamefont {Ruan},\ and\ \citenamefont {Fan}}]{ZhuTengfeng2017}%
  \BibitemOpen
  \bibfield  {author} {\bibinfo {author} {\bibfnamefont {T.}~\bibnamefont
  {Zhu}}, \bibinfo {author} {\bibfnamefont {Y.}~\bibnamefont {Zhou}}, \bibinfo
  {author} {\bibfnamefont {Y.}~\bibnamefont {Lou}}, \bibinfo {author}
  {\bibfnamefont {H.}~\bibnamefont {Ye}}, \bibinfo {author} {\bibfnamefont
  {M.}~\bibnamefont {Qiu}}, \bibinfo {author} {\bibfnamefont {Z.}~\bibnamefont
  {Ruan}}, \ and\ \bibinfo {author} {\bibfnamefont {S.}~\bibnamefont {Fan}},\
  }\href@noop {} {\bibfield  {journal} {\bibinfo  {journal} {Nature
  Communications}\ }\textbf {\bibinfo {volume} {8}},\ \bibinfo {pages} {15391}
  (\bibinfo {year} {2017})}\BibitemShut {NoStop}%
\bibitem [{\citenamefont {Li}\ \emph {et~al.}(2009)\citenamefont {Li},
  \citenamefont {Zhu}, \citenamefont {Cai}, \citenamefont {Borysiak},
  \citenamefont {Han}, \citenamefont {Chen}, \citenamefont {Piner},
  \citenamefont {Colombo},\ and\ \citenamefont {Ruoff}}]{LiXuesong2009}%
  \BibitemOpen
  \bibfield  {author} {\bibinfo {author} {\bibfnamefont {X.}~\bibnamefont
  {Li}}, \bibinfo {author} {\bibfnamefont {Y.}~\bibnamefont {Zhu}}, \bibinfo
  {author} {\bibfnamefont {W.}~\bibnamefont {Cai}}, \bibinfo {author}
  {\bibfnamefont {M.}~\bibnamefont {Borysiak}}, \bibinfo {author}
  {\bibfnamefont {B.}~\bibnamefont {Han}}, \bibinfo {author} {\bibfnamefont
  {D.}~\bibnamefont {Chen}}, \bibinfo {author} {\bibfnamefont {R.~D.}\
  \bibnamefont {Piner}}, \bibinfo {author} {\bibfnamefont {L.}~\bibnamefont
  {Colombo}}, \ and\ \bibinfo {author} {\bibfnamefont {R.~S.}\ \bibnamefont
  {Ruoff}},\ }\href@noop {} {\bibfield  {journal} {\bibinfo  {journal} {Nano
  Letters}\ }\textbf {\bibinfo {volume} {9}},\ \bibinfo {pages} {4359}
  (\bibinfo {year} {2009})}\BibitemShut {NoStop}%
\bibitem [{\citenamefont {Zhu}\ \emph {et~al.}(2013)\citenamefont {Zhu},
  \citenamefont {Yan}, \citenamefont {Jepsen}, \citenamefont {Hansen},
  \citenamefont {Mortensen},\ and\ \citenamefont {Xiao}}]{ZhuXiaolong2013}%
  \BibitemOpen
  \bibfield  {author} {\bibinfo {author} {\bibfnamefont {X.}~\bibnamefont
  {Zhu}}, \bibinfo {author} {\bibfnamefont {W.}~\bibnamefont {Yan}}, \bibinfo
  {author} {\bibfnamefont {P.~U.}\ \bibnamefont {Jepsen}}, \bibinfo {author}
  {\bibfnamefont {O.}~\bibnamefont {Hansen}}, \bibinfo {author} {\bibfnamefont
  {N.~A.}\ \bibnamefont {Mortensen}}, \ and\ \bibinfo {author} {\bibfnamefont
  {S.}~\bibnamefont {Xiao}},\ }\href@noop {} {\bibfield  {journal} {\bibinfo
  {journal} {Applied Physics Letters}\ }\textbf {\bibinfo {volume} {102}},\
  \bibinfo {pages} {131101} (\bibinfo {year} {2013})}\BibitemShut {NoStop}%
\bibitem [{\citenamefont {Garcia~de Abajo}(2014)}]{GarciadeAbajo2014}%
  \BibitemOpen
  \bibfield  {author} {\bibinfo {author} {\bibfnamefont {F.~J.}\ \bibnamefont
  {Garcia~de Abajo}},\ }\href@noop {} {\bibfield  {journal} {\bibinfo
  {journal} {ACS Photonics}\ }\textbf {\bibinfo {volume} {1}},\ \bibinfo
  {pages} {135} (\bibinfo {year} {2014})}\BibitemShut {NoStop}%
\bibitem [{\citenamefont {Gao}\ \emph {et~al.}(2012)\citenamefont {Gao},
  \citenamefont {Shu}, \citenamefont {Qiu},\ and\ \citenamefont
  {Xu}}]{GaoWeilu2012}%
  \BibitemOpen
  \bibfield  {author} {\bibinfo {author} {\bibfnamefont {W.}~\bibnamefont
  {Gao}}, \bibinfo {author} {\bibfnamefont {J.}~\bibnamefont {Shu}}, \bibinfo
  {author} {\bibfnamefont {C.}~\bibnamefont {Qiu}}, \ and\ \bibinfo {author}
  {\bibfnamefont {Q.}~\bibnamefont {Xu}},\ }\href@noop {} {\bibfield  {journal}
  {\bibinfo  {journal} {ACS Nano}\ }\textbf {\bibinfo {volume} {6}},\ \bibinfo
  {pages} {7806} (\bibinfo {year} {2012})}\BibitemShut {NoStop}%
\bibitem [{\citenamefont {Lou}\ \emph {et~al.}(2016)\citenamefont {Lou},
  \citenamefont {Pan}, \citenamefont {Zhu},\ and\ \citenamefont
  {Ruan}}]{LouPanZhuRuan16}%
  \BibitemOpen
  \bibfield  {author} {\bibinfo {author} {\bibfnamefont {Y.}~\bibnamefont
  {Lou}}, \bibinfo {author} {\bibfnamefont {H.}~\bibnamefont {Pan}}, \bibinfo
  {author} {\bibfnamefont {T.}~\bibnamefont {Zhu}}, \ and\ \bibinfo {author}
  {\bibfnamefont {Z.}~\bibnamefont {Ruan}},\ }\href@noop {} {\bibfield
  {journal} {\bibinfo  {journal} {Journal of the Optical Society of America B}\
  }\textbf {\bibinfo {volume} {33}},\ \bibinfo {pages} {819} (\bibinfo {year}
  {2016})}\BibitemShut {NoStop}%
\bibitem [{\citenamefont {Ruan}\ \emph {et~al.}(2014)\citenamefont {Ruan},
  \citenamefont {Wu}, \citenamefont {Qiu},\ and\ \citenamefont
  {Fan}}]{Ruan2014Spatial}%
  \BibitemOpen
  \bibfield  {author} {\bibinfo {author} {\bibfnamefont {Z.}~\bibnamefont
  {Ruan}}, \bibinfo {author} {\bibfnamefont {H.}~\bibnamefont {Wu}}, \bibinfo
  {author} {\bibfnamefont {M.}~\bibnamefont {Qiu}}, \ and\ \bibinfo {author}
  {\bibfnamefont {S.}~\bibnamefont {Fan}},\ }\href@noop {} {\bibfield
  {journal} {\bibinfo  {journal} {Optics Letters}\ }\textbf {\bibinfo {volume}
  {39}},\ \bibinfo {pages} {3587} (\bibinfo {year} {2014})}\BibitemShut
  {NoStop}%
\bibitem [{\citenamefont {Ruan}(2015)}]{ruan2015spatial}%
  \BibitemOpen
  \bibfield  {author} {\bibinfo {author} {\bibfnamefont {Z.}~\bibnamefont
  {Ruan}},\ }\href@noop {} {\bibfield  {journal} {\bibinfo  {journal} {Optics
  Letters}\ }\textbf {\bibinfo {volume} {40}},\ \bibinfo {pages} {601}
  (\bibinfo {year} {2015})}\BibitemShut {NoStop}%
\bibitem [{\citenamefont {Haus}(1984)}]{haus1984waves}%
  \BibitemOpen
  \bibfield  {author} {\bibinfo {author} {\bibfnamefont {H.}~\bibnamefont
  {Haus}},\ }\href@noop {} {\emph {\bibinfo {title} {{Waves and fields in
  optoelectronics.}}}}\ (\bibinfo  {publisher} {Prentice-Hall},\ \bibinfo
  {year} {1984})\BibitemShut {NoStop}%
\bibitem [{\citenamefont {Fan}\ \emph {et~al.}(2003)\citenamefont {Fan},
  \citenamefont {Suh},\ and\ \citenamefont {Joannopoulos}}]{fan2003temporal}%
  \BibitemOpen
  \bibfield  {author} {\bibinfo {author} {\bibfnamefont {S.}~\bibnamefont
  {Fan}}, \bibinfo {author} {\bibfnamefont {W.}~\bibnamefont {Suh}}, \ and\
  \bibinfo {author} {\bibfnamefont {J.~D.}\ \bibnamefont {Joannopoulos}},\
  }\href@noop {} {\bibfield  {journal} {\bibinfo  {journal} {Journal of the
  Optical Society of America A}\ }\textbf {\bibinfo {volume} {20}},\ \bibinfo
  {pages} {569} (\bibinfo {year} {2003})}\BibitemShut {NoStop}%
\end{thebibliography}
\end{document}